\documentclass[ reprint,nofootinbib,amsmath,amssymb,aps]{revtex4-1}
\usepackage{graphicx}
\usepackage{dcolumn}
\usepackage[colorlinks,linkcolor=magenta,anchorcolor=cyan,citecolor=blue]{hyperref}
\usepackage{bm}
\usepackage[multiple]{footmisc}
\newcommand{\fig}[1]{Fig.\ref{#1}}
\def\be{\begin{equation}}
\def\ee{\end{equation}}
\def\ba{\begin{eqnarray}}
\def\ea{\end{eqnarray}}

\def\nn{\nonumber}
\def\lf{\left}
\def\rt{\right}

\newcommand{\eq}[1]{(\ref{#1})}

\def\dbar{{\mathchar '26\mkern -10mu\delta}}

\def\nn{\nonumber}\def\lf{\left}\def\rt{\right}\def\q{\theta} \def\w{\omega}\def\r {\rho}  \def\y {\psi}   \def\p {\pi} \def\a {\alpha} \def\s {\sigma} \def\d {\delta} \def\f {\phi}  \def\h {\eta}  \def\k {\kappa} \def\l {\lambda}    \def\b {\beta}  \def\m {\mu} \def\pd {\partial}   \def \e { \varepsilon}
\def\Q{\Theta} \def\W{\Omega} \def\Y {\Psi}    \def\S {\Sigma}   \def\G {\Gamma} \def\L {\Lambda}    \def\grad{\nabla}\def\.{\cdot}
\def\math {\mathcal}
\begin{document}
\title{ Surface term, corner term, and action growth in $F($Riemann$)$ gravity theory}
\author{Jie Jiang$^1$}
\email{jiejiang@mail.bnu.edu.cn}
\author{Hongbao Zhang$^{1,2}$}
\email{hzhang@vub.ac.be}
\affiliation{$^1$Department of Physics, Beijing Normal University, Beijing 100875, China\label{addr1}}
\affiliation{$^2$Theoretische Natuurkunde, Vrije Universiteit Brussel,
 and The International Solvay Institutes, Pleinlaan 2, B-1050 Brussels, Belgium\label{addr2}}
\date{\today}

\begin{abstract}
After reformulating $F($Riemann$)$ gravity theory as a second derivative theory by introducing two auxiliary fields to the bulk action, we derive the surface term as well as the corner term supplemented to the bulk action for a generic non-smooth boundary such that the variational principle is well posed. We also introduce the counter term to make the boundary term invariant under the reparametrization for the null segment. Then as a demonstration of the power of our formalism, not only do we apply our expression for the full action to evaluate the corresponding action growth rate of the Wheeler-DeWitt patch in the Schwarzchild anti-de Sitter black hole for the $F(R)$ gravity and critical gravity, where the corresponding late time behavior recovers the previous one derived by other approaches, but also in the asymptotically Anti-de Sitter black hole for the critical Einsteinian cubic gravity, where the late time growth rate vanishes but still saturates the Lloyd bound.
\end{abstract}
\maketitle
\section{Introduction}
Generically, in order to make the variational principle well posed for gravity theories, one is required to add the surface term to the bulk action. In this way, the Gibbons-Hawking-York (GHY) surface term is introduced for the case of Einstein gravity, but is applicable only to a non-null boundary\cite{Gowdy,York,GH}. For a null boundary, the surface term has also been investigated recently\cite{Neiman,PCMP,PCP,LMPS,HF}. Moreover, if the boundary is non-smooth, $i.e.$, the boundary contains some corners intersected by the segments, the additional corner term has to be added to the action\cite{Hayward,JSSS}. On the other hand, although the non-null surface terms have been developed for other gravitational theories, such as $F(R)$ gravity\cite{CDM,GCT}, Gauss-Bonnet gravity\cite{Bunch,Myers}, Lanczos-Lovelock theory\cite{PK,BCLR,CPP,Cano}, and other higher derivative theories\cite{ST,TTEM,BCR}, the corresponding null surface term has not been fully explored.

However, for a generic higher order gravitational theory as usually formulated, due to higher-derivative terms, it is hard to find an appropriate surface term to make the variational principle well posed\cite{DKH}. But at least for $F($Riemann$)$ gravity, this problem can be circumvented by introducing two auxiliary fields, because this allows us to recast the action as a second order gravitational theory, which is on-shell equivalent to the original action\cite{DMY}. Furthermore, if the auxiliary fields on the boundary can be shown by the Hamiltonian analysis to be independent of the extrinsic curvature\footnote{It is noteworthy that Lanczos-Lovelock theory does not satisfy this requirement and will not be treated in this paper. Readers are referred to \cite{Cano,CP} for this theory.}, then for a smooth non-null boundary a generalized GHY term can be found to establish the well posed variational principle. In this paper, we shall focus exclusively on this situation and formulate the well posed variational principle for more general circumstances, where the boundary is not necessarily required to be non-null or smooth.

Another motivation to evaluate the full action with a non-smooth boundary including null segments comes from the ``complexity equals action" (CA) conjecture\cite{BRSSZ1,BRSSZ2}. This conjecture states that the complexity of a particular state $|\y(t_L,t_R)\rangle$ on the boundary is given by
\ba
\math{C}\lf(|\y(t_L,t_R)\rangle\rt)\equiv\frac{I}{\p\hbar}\,,
\ea
where $I$ is the on-shell action in the corresponding Wheeler-DeWitt (WDW) patch, enclosed by the past and future light sheets sent into the bulk spacetime from the boundary time slices $t_L$ and $t_R$. As an application of our formulation of the full action for $F($Riemann$)$ gravity, we shall evaluate the action growth rate of the WDW patch in the Schwarzschild anti-de Sitter (SAdS) black hole for both the $F(R)$ gravity and critical gravity. This thus makes up the deficiency of the approaches developed in \cite{BRSSZ1,BRSSZ2}, which can only give rise to the late time behavior of the action growth rate\cite{AFNV,GWLL}. To further demonstrate the power of our formalism, we also evaluate the action growth rate of the WDW patch in the asymptotically AdS black hole for the critical Einsteinian cubic gravity. The resulting late time growth rate still saturates the Lloyd bound although vanishes.

This paper is structured as follows. In Sec.\ref{se2}, we follow the strategy developed in \cite{DMY} to introduce the two auxiliary fields to reformulate the original action and evaluate its variation. After this, we derive the required boundary term to make the variational principle well posed for both non-null segments and null segments of a non-smooth boundary in Sec.\ref{se3} and Sec.\ref{se4}, respectively. As an application of the resulting full action, Section \ref{se5} devotes an explicit calculation of the action growth rate for the WDW patch in the SAdS black hole for both $F(R)$ gravity and critical gravity, as well as in the asymptotically AdS black hole for the critical Einsteinian cubic gravity. We conclude our paper in Sec. \ref{se6}.

\section{Reformulation of $F($Riemann$)$ gravity theory}\label{se2}
The conventional bulk action for $F($Riemann$)$ gravity is given by
\ba\label{acb}
I_{\text{bulk}}=\int_{\math{M}}d^{d+2}x\sqrt{-g}F(R_{abcd},g_{ab})\,
\ea
with $F$ an arbitrary function of $R_{abcd}$ and $g_{ab}$.
Its variation can be obtained as
\ba\label{action0}
\d I_{\text{bulk}}=\int_{\math{M}}d^{d+2}x\sqrt{-g}E_{ab}\d g^{ab}+\int_{\pd \math{M}}\dbar v^a d\S_a\, .
\ea
Here $d\S_a$ is the outward-directed surface element on $\pd \math{M}$, and
\ba\label{dv0}
\dbar v^c&=&2 P_a{}^{bcd}\d{\G^a}_{bd}+2\d g_{bd}\grad_aP^{abcd}\,
\ea
with $P^{abcd}=\frac{\pd F}{\pd R_{abcd}}$. In addition, the symbol $\dbar$  indicates an infinitesimal quantity which can not be written as the variation of any quantity. Obviously, $E_{ab}=0$ is simply the equation of motion. But in order to give rise to a well posed variational principle, we must supplement a boundary term $I_{\text{bdry}}$ such that
\ba
\d I_{\text{bdry}} = -\int_{\pd \math{M}} \dbar v^ad\S_a + \int_{\pd \math{M}} p_N \d q^N d\S\,
\ea
with $q^N$ the intrinsic geometric quantity as well as its derivatives to the boundary. If the boundary is smooth, the boundary term involves only the surface term $I_{\text{surf}}$. On the other hand, if the boundary is non-smooth, not only does the boundary term include the surface term, but also the corner term $I_{\text{corner}}$.

 However, it is generically difficult to find the corresponding boundary term, if any, for the bulk action (\ref{acb}). Gratefully this problem can be circumvented by introducing two auxiliary fields $\y_{abcd}$ and $\f_{abcd}$, which allows us to  recast the original bulk action (\ref{acb}) into the following form\cite{DMY}
\ba\label{action2}\begin{aligned}
I_{\text{bulk}}=&\int_{\math{M}}d^{d+2}x\sqrt{-g}\times\\
&\lf[F(\f_{abcd},g_{ab})-\y^{abcd}\lf(\f_{abcd}-R_{abcd}\rt)\rt]\,,
\end{aligned}\ea
where we demand these two auxiliary fields have the same symmetries as $R_{abcd}$. The variation of this new action can be expressed as
\ba\label{action}\begin{aligned}
\d I_{\text{bulk}}=&\int_{\math{M}}d^{d+2}x\sqrt{-g}\times\\
&\lf(E_{ab}\d g^{ab}+E_\f^{abcd}\d \f_{abcd}+E_{\y}^{abcd}\d \y_{abcd}\rt)\\
&+\int_{\pd \math{M}}\dbar v^ad\S_a\,
\end{aligned}\ea
with
\ba\label{em}\begin{aligned}
E_{\f}^{abcd}&=\frac{\pd F(\f_{abcd},g_{ab})}{\pd \f_{abcd}}- \y^{abcd}\,,\\
E_{\y}^{abcd}&=R^{abcd} -\f^{abcd}\,,
\end{aligned}\ea
and
\ba\label{dv}
\dbar v^c=2 {{\y}_a}{}^{bcd}\d{\G^a}_{bd}+2\d g_{bd}\grad_a{\y}^{abcd}\,.
\ea
 With the equations of motion ${E}_{\y}^{abcd}=0$ and $ E_\f^{abcd}=0$ satisfied, this new action is equivalent to the original one. In particular, the corresponding boundary term is identified by the Hamiltonian analysis in \cite{DMY} for the smooth non-null boundary.

In what follows, we shall derive the boundary term for a more general boundary by requiring this new action have a well posed variational principle.

\section{Non-null segments}\label{se3}
\subsection{Variation of geometric quantities}\label{se31}
We first present the variation of geometric quantities associated with the segment of the boundary, which is either spacelike or timelike. To achieve this, we choose the gauge in which the segment under consideration is fixed when we perform the variation. With this in mind, we have the variation of the outward-directed normal vector
\ba
\d n_a=\d a\, n_a\,
\ea
with $\d a=-\frac{\e}{2}\d g^{ab}n_an_b$, where $\e=n_an^a$. Whence we further have
\ba
\d n^a=-\d a\,n^a-\e\dbar A^a\,
\ea
with $\dbar A^a=-\e h^a{}_b\d g^{bc}n_c$, where the induced metric is given by
\ba\label{htog}
h^{ab} = g^{ab}-\e\, n^an^b\,,
\ea
which is tangent to the segment.
The variation of the metric can be further expressed as
\ba\label{dg}
\d g^{ab}=-2\e\d a\, n^a n^b-\dbar A^a n^b-\dbar A^b n^a+\d h^{ab}\,,
\ea
whereby it is not hard to show
\ba \label{once}
h^a{}^{d}h^b{}_{e}h^c{}_f\grad_d \d g^{ef}=D^a \d h^{bc}- K^{ac}\dbar A^b-K^{ab}\dbar A^c
\ea
with $K^{ab}=h^{ac}h^{bd}\nabla_cn_d$ the extrinsic curvature.

Finally, for the later calculations, we would like to present two expressions for the variation of the extrinsic curvature. The first one is given by
\ba\label{K1}\begin{aligned}
\d K^{ab} =& \d \lf(h^{ac}h^{bd}\grad_c n_d\rt)\\
=& \d h^{ac} h^{bd}\grad_c n_d+ h^{ac} \d h^{bd}\grad_c n_d \\
&- h^{ac} h^{bd} n_e{\d\G^e}_{cd}+h^{ac} h^{bd}\grad_c \d n_d\\
=& \d h^{ac} {K^b}_c+ \d h^{bc} {K^a}_c +\d a\,K^{ab} - h^{ac} h^{bd} n_e{\d\G^e}_{cd}\,,
\end{aligned}\ea
and the second one is given by
\ba\label{K2}\begin{aligned}
\d K^{ab} =& \d \lf(h^{ac}{h^b}_d\grad_c n^d\rt)\\
=& \d h^{ac} {h^b}_d\grad_c n^d+ h^{ac} \d {h^b}_d\grad_c n^d\\
&+h^{ac} {h^b}_d{\d\G^d}_{ce}n^e +h^{ac} {h^b}_d\grad_c \d n^d\\
=& \d h^{ac} {K^b}_c -\d a\,K^{ab} -\e D^a \dbar A^b +h^{ac} {h^b}_d{\d\G^d}_{ce}n^e\,,
\end{aligned}\ea
where we have used $\d {h^b}_d=\d\lf( {\d^b}_d-\e n^b n_d\rt)=-\dbar A^b n_d$ and $D_a$ as the covariant derivative operator of the induced metric.

\subsection{Surface term on the boundary}\label{se32}
As to the spacelike/timelike segment of the boundary, the boundary term in the variation of the bulk action (\ref{dg}) can be written as
\ba\label{bdry1}\begin{aligned}
&\int_{\S}\dbar v^a d\S_a=\e\int_{\S}n_a\dbar v^a d\S\\
&=\e\int_{\S}\lf[ 2n_c{{\y}_a}{}^{bcd}\d{\G^a}_{bd}+2n_c\d g_{bd}\grad_a{\y}^{abcd}\rt]d\S.
\end{aligned}\ea
The first term in \eq{bdry1} can be further evaluated as
\ba\label{1}\begin{aligned}
&2n_c{{\y}_a}{}^{bcd}\d{\G^a}_{bd}=2n_c{{\y}_a}{}^{bcd}\d^a{}_{a_1}\d^{b_1}{}_{b}\d^{d_1}{}_{d}\d{\G^{a_1}}_{b_1d_1}\\
&=2n_c{{\y}_a}{}^{bcd}\lf(\e n^an_{a_1}+h^a{}_{a_1}\rt)\lf(\e n^{b_1}n_b+h^{b_1}{}_{b}\rt)\\
&\times\lf(\e n^{d_1}n_d+h^{d_1}{}_{d}\rt)\d{\G^{a_1}}_{b_1d_1}\\
&=-2\e\Psi _{ab}\lf( h^{\text{be}} h^a{}_d \delta \Gamma ^d{}_{{ce}} n^c- h^{{ad}} h^{{be}}  \delta \Gamma ^c{}_{{de}}n_c\rt)\\
&+2n^d h^g{}_c h^{ea}h^{fb} \y_{{gedf}}\delta \Gamma^c{}_{ab}\,,
\end{aligned}\ea
where we have used the symmetries of the auxiliary field $\y_{abcd}$ and the definition
\ba\label{Yab}
\Y_{ab}\equiv\y_{acbd}n^cn^d\, .
\ea
Substituting \eq{K1} and \eq{K2} into the first two terms in \eq{1}, we end up with
\ba
\begin{aligned}
&-2\e\Psi _{ab}\lf( h^{be} h^a{}_d \delta \Gamma ^d{}_{{ce}} n^c- h^{{ad}} h^{{be}}  \delta \Gamma ^c{}_{{de}}n_c\rt)\\
&=-2\e\Y_{ab}\lf(2\d K^{ab}-3 K^a{}_c \d h^{cb}+ \e D^a\dbar A^b\rt)\,,
\end{aligned}\ea
where the property $\Y_{ab}=\Y_{ba}$ has been used. For the third term in \eq{1}, we have
\ba\begin{aligned}
&2n^d h^g{}_c h^{ea}h^{fb} \y_{{gedf}}\delta \Gamma^c{}_{ab}\\
&=n^d h^{ea}h^{fb} \y_{{gedf}}h^{gc}\lf(\grad_a \d g_{cb}+\grad_b \d g_{ca}-\grad_c \d g_{ab}\rt)\\
&=2n^d  \y_{gedf}h^{ea}h^{fb}h^{gc}\grad_a \d g_{cb}\\
&=2n^d \y_{gedf} \lf(K^{eg}\dbar A^f+K^{ef}\dbar A^g-D^e\d h^{fg}\rt)\\
&=2n^d \y_{cadb} \lf(K^{ab}\dbar A^c-D^a\d h^{bc}\rt)\,,\\
\end{aligned}\ea
where \eq{once} as well as $\d g_{cb}=-g_{ca}g_{bd}\d g^{ab}$ has been used in the second step.
Then \eq{1} reduces to
\ba\label{1term}
2n_c{{\y}_a}{}^{bcd}&\d{\G^a}_{bd}= -2\e\Y_{ab}\lf(2\d K^{ab}-3 K^a{}_c \d h^{cb}+ \e D^a\dbar A^b\rt)\nn\\
&+2n^d \y_{cadb} \lf(K^{ab}\dbar A^c-D^a\d h^{bc}\rt)\,.
\ea
On the other hand, the second term in \eq{bdry1} can be expressed as
\ba\label{2term}\begin{aligned}
&2n_c\d g_{bd}\grad_a{\y}^{abcd}\\
&=-2n^an^b\dbar A^ch^{ef}\grad_f \y_{beac}+2n^a\delta h^{bc}\nabla ^e \y_{beac}\\
&=-2\dbar A^a D^b \Y_{ab}+2 \dbar A^a \y_{acbd}K^{bc}n^d+2n^a\delta h^{bc}\nabla ^e \y_{beac}\,.
\end{aligned}\ea
Plugging \eq{1term} and \eq{2term} into \eq{bdry1}, we have
\ba\label{bj2}\begin{aligned}
&n_a\dbar v^a= -2\e\Y_{ab}\lf(2\d K^{ab}-3 K^a{}_c \d h^{cb}+ \e D^a\dbar A^b\rt)\\
&+2n^d \y_{cadb} \lf(K^{ab}\dbar A^c-D^a\d h^{bc}\rt)-2\dbar A^a D^b \Y_{ab}\\
&+2 \dbar A^a \y_{acbd}K^{bc}n^d+2n^a\delta h^{bc}\nabla ^e \y_{beac}\\
&=-4\e\Y_{ab}\d K^{ab}-2D^a\lf(\dbar A^b\Y_{ab}\rt)\\
&+\lf(2n^a \nabla ^e \y_{beac}+6\e\Y_{ab} K^a{}_c \rt)\d h^{bc}-2n^d \y_{cadb}D^a\d h^{bc}\,.
\end{aligned}\ea
Now by requiring both $\d h^{ab}$ and $\d \Y_{ab}$ vanish on the boundary, we have
\ba\label{boundary1}
\int_{\S}\dbar v^a d\S_a=-4\d\lf( \int_\S \Y_{ab} K^{ab}d\S\rt)-2\e \int_{\pd \S}\dbar A^b\Y_{ba}dS^a\nn\,.\\
\ea
If the boundary is smooth, $\pd \S=\pd^2\math{M}=0$ implies that the second term vanishes. Accordingly, the bulk action can be supplemented with the surface term
\ba\label{Isurf}
I_{\text{surf}}=4 \int_{\pd \math{M}} \Y_{ab} K^{ab}d\S
\ea
to make the variational principle well posed. However, if the boundary is non-smooth, the second term does not vanish. In this case, to have a well posed variational principle, we need add the additional corner term such that
\ba\begin{aligned}
&\d I_{\text{corner}} = 2\sum_{s}\lf(\e \int_{\pd \S_s}\dbar A^b\Y_{ba}dS^a\rt)\\
&=2\sum_{s,s'}\int_{\math{C}_{ss'}}\lf(\e_s\dbar A_s^a\Y_{s ab}dS_s^b+\e_{s'}\dbar A_{s'}^a\Y_{{s'} ab}dS_{s'}^b\rt)\,.
   \end{aligned}
\ea
where the subscripts $s,s'$ denote the segments of the boundary and  $\math{C}_{ss'}=\pd \S_s\cap \pd \S_{s'}$ denotes the joint intersected by the segments $\S_s$ and $\S_{s'}$. For simplicity, we would like to define the corner term $I_{\math{C}_{ss'}}$ contributed by the joint $\math{C}_{ss'}$, which satisfies
\ba\label{corner}
\d I_{\math{C}_{ss'}}=2\int_{\math{C}_{ss'}}\lf(\e_s\dbar A_s^a\Y_{s ab}dS_s^b+\e_{s'}\dbar A_{s'}^a\Y_{{s'}ab}dS_{s'}^b\rt)\nn\,.\\
\ea
 Next, we shall separately derive the explicit expression of the corner term for all kinds of joints intersected by the segments of the boundary.

\subsection{Corner term on the boundary}\label{se33}
\subsubsection{Timelike joint}
\begin{figure}
\centering
\includegraphics[width=2.5in,height=2in]{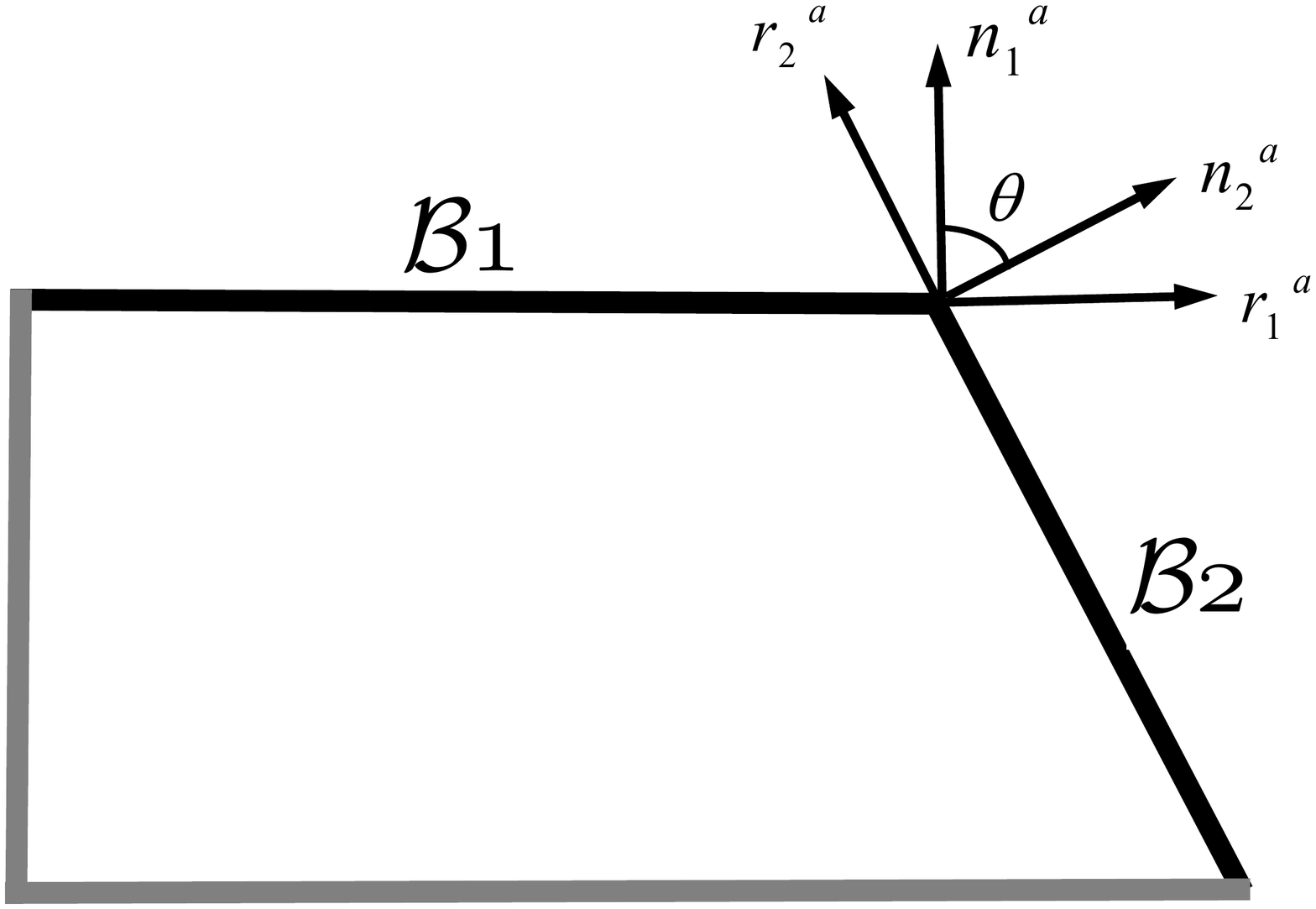}
\caption{The timelike joint is intersected by two timelike segments $\math{B}_1$ and $\math{B}_2$.}
\label{tlj}
\end{figure}

 As depicted in \fig{tlj}, we first consider the timelike joint $\math{C}$ intersected by two timelike segments of the boundary $\math{B}_1$ and $\math{B}_2$, $i.e.$, $\math{C}=\math{B}_1\cap\math{B}_2$. Note that the condition $\d h_s^{ab}=0$, we have
 \ba\begin{aligned}
 \d g^{ab}&=&-2\d a_1\, n_1^a n_1^b-\dbar A_1^a n_1^b-\dbar A_1^b n_1^a\\
&=&-2\d a_2\, n_2^a n_2^b-\dbar A_2^a n_2^b-\dbar A_2^b n_2^a\label{g2}
 \end{aligned}\ea
 at the joint $\math{C}$. In addition, for each normal vector $n_{s a}$ at the joint $\math{C}$, there exists another normal vector $r_{s a}$ to the joint, which points outwards from $\math{B}_s$ and satisfies $r_{s}\cdot n_{s}=0$. $\{n_s^a,r_s^a\}$ forms a pair of unit normals at the joint $\math{C}$, and the two pairs can be related to each other by a rotation
\ba\label{trsf}\begin{aligned}
n_2^a&=n_1^a \cos \q+r_1^a \sin \q,\\
r_2^a&=n_1^a \sin \q-r_1^a \cos \q
\end{aligned}\ea
for some rotation parameter $\q$.
Substitute \eq{trsf} into \eq{g2} and make a decomposition $\dbar A_s^a=\dbar A_s^r r_s^a+\dbar \hat{A}_s^a$ with $\dbar \hat{A}_s^a$ a tangent vector of the joint $\math{C}$, then we have
\ba\label{crdg}\begin{aligned}
&-2\d a_2\, n_2^a n_2^b-\dbar A_2^r r_2^a n_2^b-\dbar A_2^r r_2^b n_2^a-\dbar \hat{A}_2^a n_2^b-\dbar \hat{A}_2^b n_2^a\\
&=-2 \cos\q\lf( \sin\q \dbar A_2^r+\cos\q \d a_2 \rt)n_1^an_1^b\\
&+\lf(\cos 2\q \dbar A_2^r-\sin2\q \d a_2\rt)(n_1^a r_1^b+n_1^b r_1^a)\\
&+2\sin\q\lf(\cos\q \dbar A_2^r-\sin\q\d a_2 \rt)r_1^a r_1^b\\
&-\sin\q  \lf(\dbar\hat{A}_2^a r_1^b+\dbar\hat{A}_2^b r_1^a\rt)-\cos\q  \lf(\dbar\hat{A}_2^a n_1^b+\dbar\hat{A}_2^b n_1^a\rt)\\
&=-2\d a_1\, n_1^a n_1^b-\d A_1^r r_1^a n_1^b-\d A_1^r r_1^b n_1^a-\d \hat{A}_1^a n_1^b-\d \hat{A}_1^b n_1^a\,,
\end{aligned}\ea
which gives rise to
\ba\label{aaa}
\d a_1 &=& \d a_2 \equiv \d a\,,\\
\dbar A_s^a &=& \tan \q\, \d a\, r_s^a\,.
\ea
On the other hand, from the transformation (\ref{trsf}), we can obtain
\ba
\cos \q=n_2\cdot n_{1}\,,
\ea
the variation of which yields
\ba\begin{aligned}
-\sin\q\, \d \q&=-\d a\, n_2^an_{1a}- \dbar A_2^an_{1a}+\d an_2^a n_{1a}\\
&=-\tan\q \,\d a\, \sin\q\,.
\end{aligned}\ea
Whence we have
\ba
\dbar A_s^a &=&\d \q r_s^a\,.
\ea

With the above preparation, the variation of the corner term can be written as
\ba\label{bjx}\begin{aligned}
\d I_{\math{C}}&=2\int_\math{C}\lf(\dbar A_1^a\Y_{1ab}dS_1^b+\dbar A_2^a\Y_{2ab}dS_2^b\rt)\\
&=2\int_{\math{C}}\lf(\Y_{1ab}\dbar A_1^a r_1^b+\Y_{2ab}\dbar A_2^ar_2^b \rt)dS\\
&=\int_{\math{C}}\hat{\Y}\d \q dS\,,
\end{aligned}\ea
where $\hat{\Y}=4\Y_{sab}r_s^ar_s^b=\y^{abcd}\epsilon_{ab}\epsilon_{cd}$ is the Wald entropy density with the binomal defined as $\epsilon_{ab}=(n_s\wedge r_s)_{ab}=2n_{[s a}r_{s b]}$, which does not depend on the choice of pairs, namely keeps invariant under the above Lorentz transformation.

The requirements $\d \Y_{ab}=0$ and $\d r_s^a=0$ lead to $\d \hat{\Y}=0$. Accordingly, the corner term can be derived as the Wald entropy density multiplied by the rotation parameter, $i.e.$,
\ba
I_{\math{C}}=\int_{\math{C}}\hat{\Y}\, \q dS\,,
\ea
which vanishes when $\q=0$ as it is expected to be the case.

\subsubsection{Spacelike joint}\label{se34}
\begin{figure}
\centering
\includegraphics[width=0.4\textwidth]{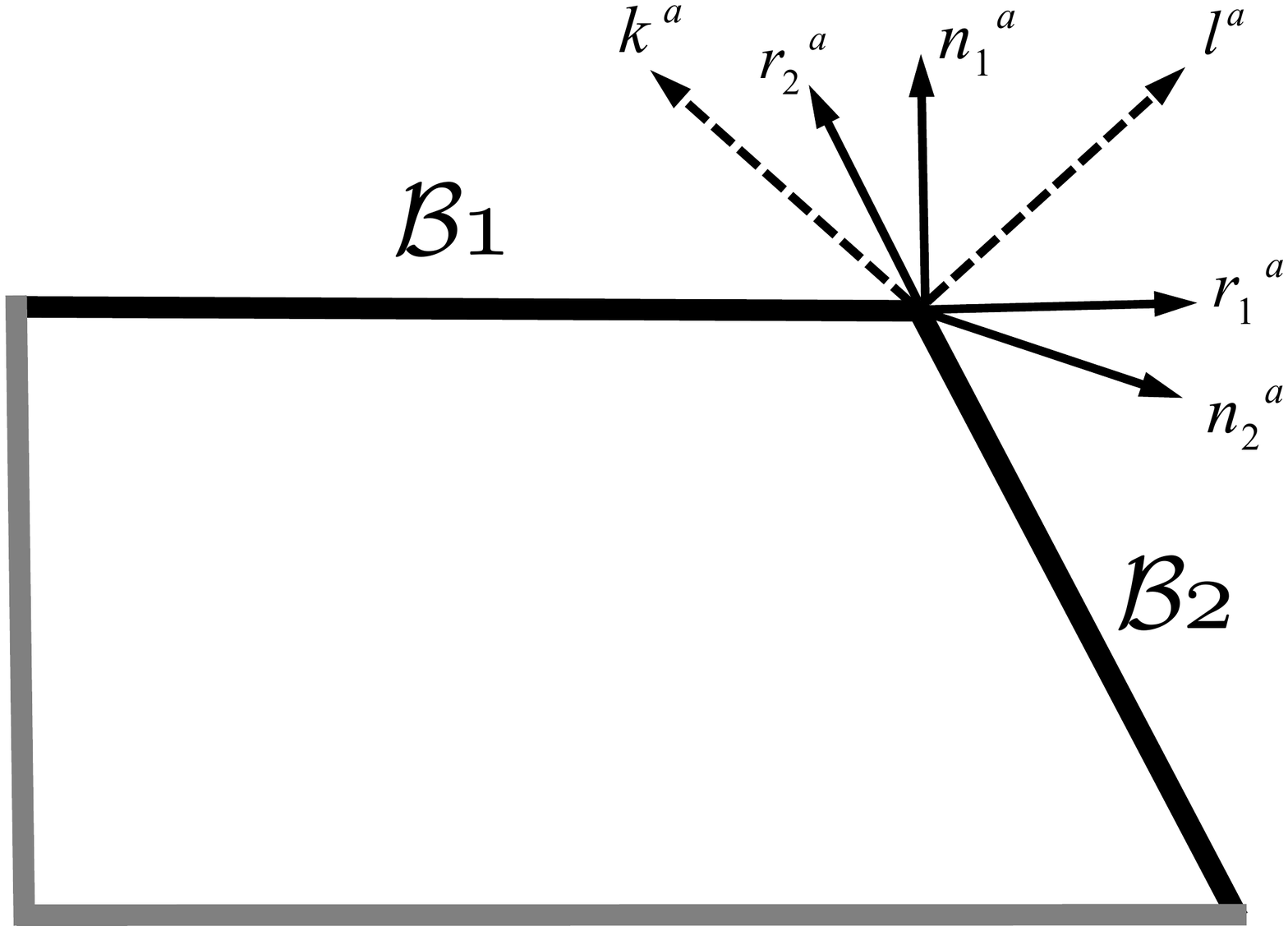}
\caption{ The spacelike joint is intersected by a spacelike segment $\math{B}_1$ and a timelike segment $\math{B}_2$.}\label{ts}
\end{figure}
 As shown in \fig{ts}, now we consider a typical type of spacelike joint $\math{C}$ intersected by a spacelike segment $\math{B}_1$ and a timelike segment $\math{B}_2$ . In this case, the two pairs of the normal vector $\{n_s^a,r_s^a \}$ can be related to each other by the boost transformation
\ba\label{trsf2}\begin{aligned}
n_2^a&=r_1^a \cosh \h-n_1^a \sinh \h\,,\\
r_2^a&=n_1^a \cosh \h-r_1^a \sinh \h\,
\end{aligned}\ea
with $\h$ the boost parameter. Substituting this transformation into the following equality
 \ba
 \d g^{ab}&=&2\d a_1\, n_1^a n_1^b-\dbar A_1^a n_1^b-\dbar A_1^b n_1^a\nonumber\\
&=&-2\d a_2\, n_2^a n_2^b-\dbar A_2^a n_2^b-\dbar A_2^b n_2^a\label{g22}
 \ea
at the joint $\math{C}$, one can show
\ba
\d a_1&=&\d a_2 \equiv \d a\,,\\
\dbar A_s^a &=& \coth \h\, \d a\, r_s^a\,.
\ea
Furthermore, by virtue of the variation of $\sinh\h = n_1\.n_{2}$, one can obtain
\ba
\dbar A_s^a=\d\h \,r_s^a\,.
\ea
Accordingly, the variation of the corresponding corner term can be expressed as
\ba\begin{aligned}
\d I_{\math{C}}&=-2\int_\math{C}\lf(\dbar A_1^a\Y_{1ab}dS_1^b-\dbar A_2^a\Y_{2ab}dS_2^b\rt)\\
&=-2\int_{\math{C}}\lf(\Y_{1ab}\dbar A_1^a r_1^b+\Y_{2ab}\dbar A_2^a r_2^b\rt)dS\\
&=-\int_{\math{C}}\hat{\Y}\d \h dS\,,\\
\end{aligned}\ea
where we have used $ dS_1^a=r_1^a dS $ and $ dS_2^a=-r_2^a dS $ due to the fact that $r_1^a$ is spacelike while $r_2^a$ is timelike. Whence one can obtain the corner term as
\ba
I_{\math{C}}=-\int_{\math{C}}\hat{\Y} \h dS\,,
\ea
where we have required the corner term satisfy the additivity rule, which will be documented in detail later on.

For the later convenience, we would like to re-express the boost parameter $\h$. To this end, as shown in \fig{ts}, we define $l^a$ to be a null vector as
\ba\begin{aligned}
l^a&=A (n_1^a+r_1^a)\\
&=B( n_2^a+ r_2^a)\,.
\end{aligned}\ea
Substitute the transformation (\ref{trsf2}) into it, then we have
\ba
B = A \lf(\cosh\h+\sinh\h\rt)=A e^{\h}\,,
\ea
which leads to a new expression for the boost parameter as
\ba\label{et1}
\h=\ln B-\ln A=\ln\lf(l\.n_2\rt)- \ln \lf(- l\. n_1\rt)\,.
\ea
By the same token, in terms of another null vector
\ba\begin{aligned}
k^a&=C ( n_1^a-r_1^a)\\
&=D ( -n_2^a+r_2^a)\,,
\end{aligned}\ea
the boost parameter can also be written as
\ba\label{et2}
\h= \ln\lf(- k\.n_1\rt)-\ln\lf(-k\.n_2\rt)\,.
\ea

\subsubsection{Other joints}\label{se35}
\begin{figure*}
\begin{minipage}[b]{.48\linewidth}
\centering
(a)
\includegraphics[width=2in,height=1.5in]{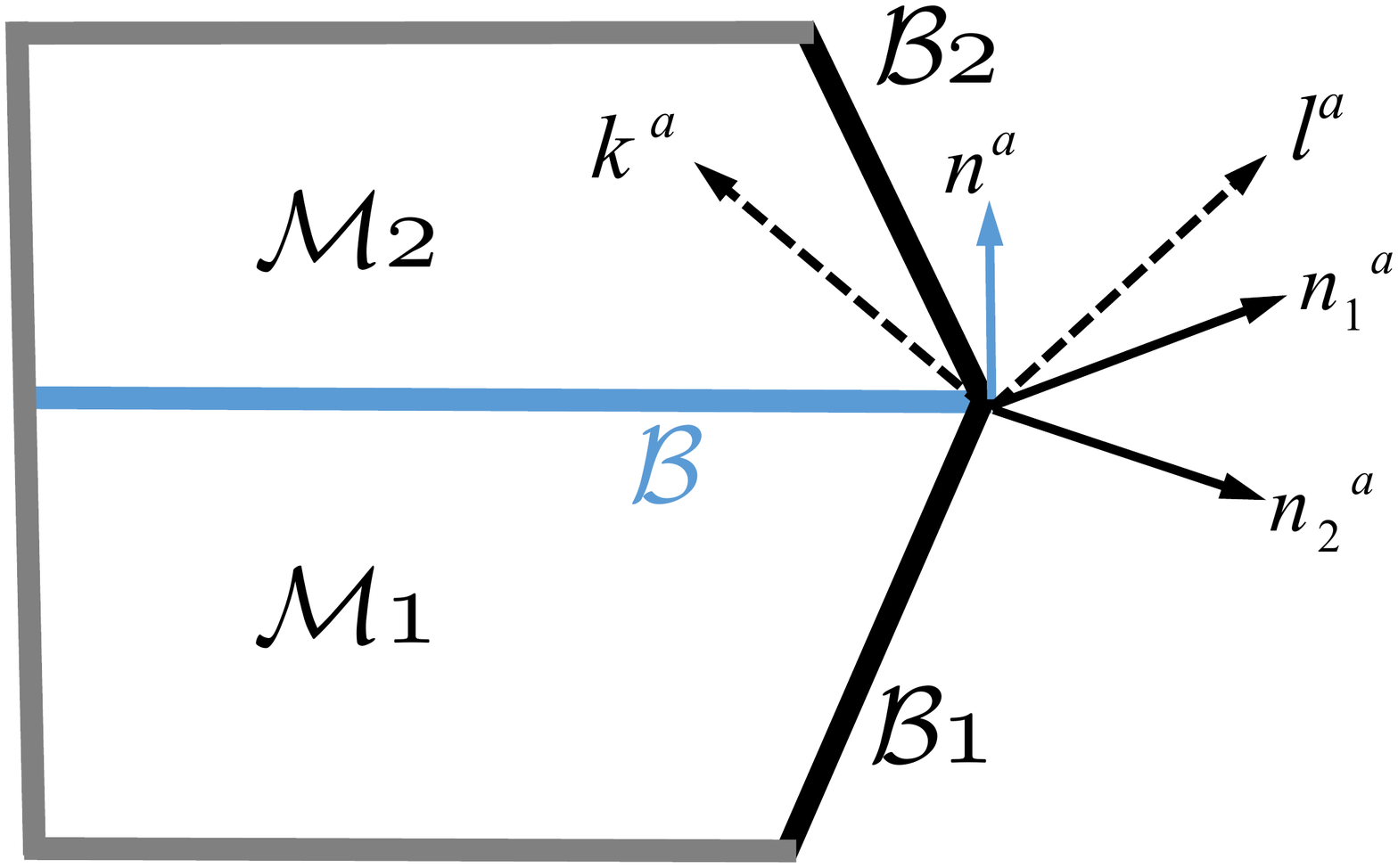}\label{a}
\end{minipage}
\begin{minipage}[b]{.48\linewidth}
\centering(b)
\includegraphics[width=2in,height=1.5in]{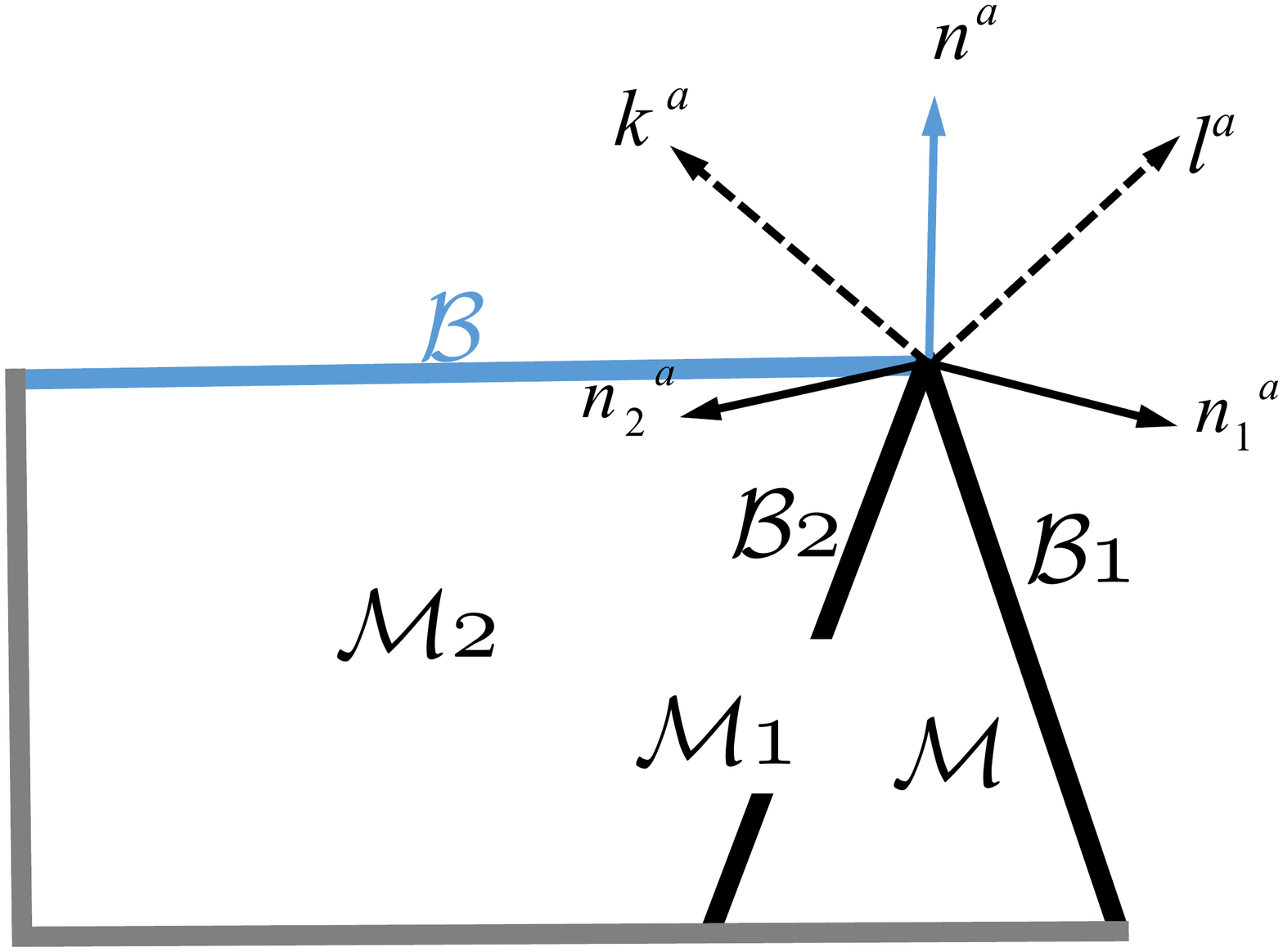}
\label{b}
\end{minipage}
\begin{minipage}[b]{.48\linewidth}
\centering(c)
\includegraphics[width=2in,height=1.5in]{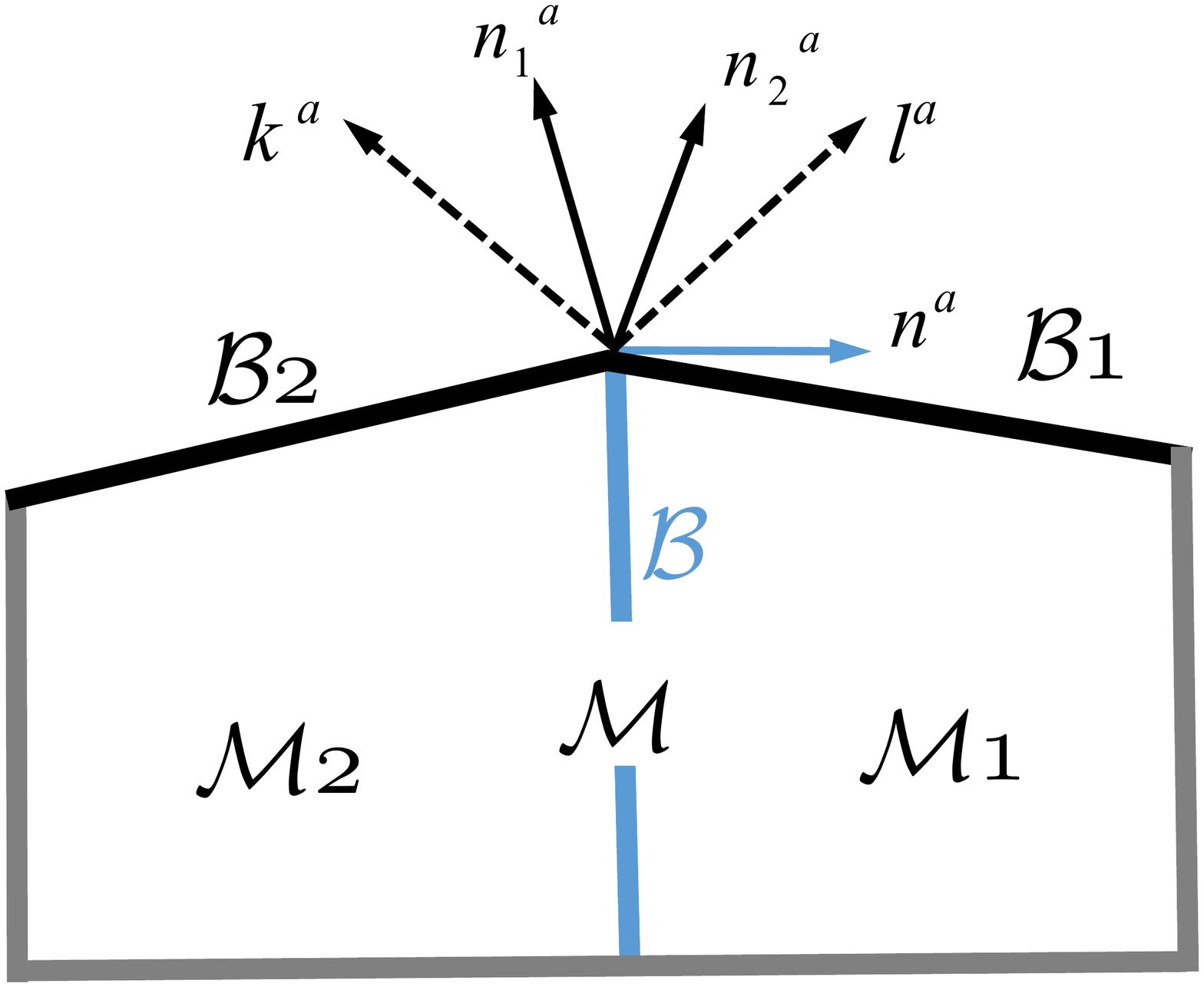}
\label{c}
\end{minipage}
\begin{minipage}[b]{.48\linewidth}
\centering(d)
\includegraphics[width=2in,height=1.5in]{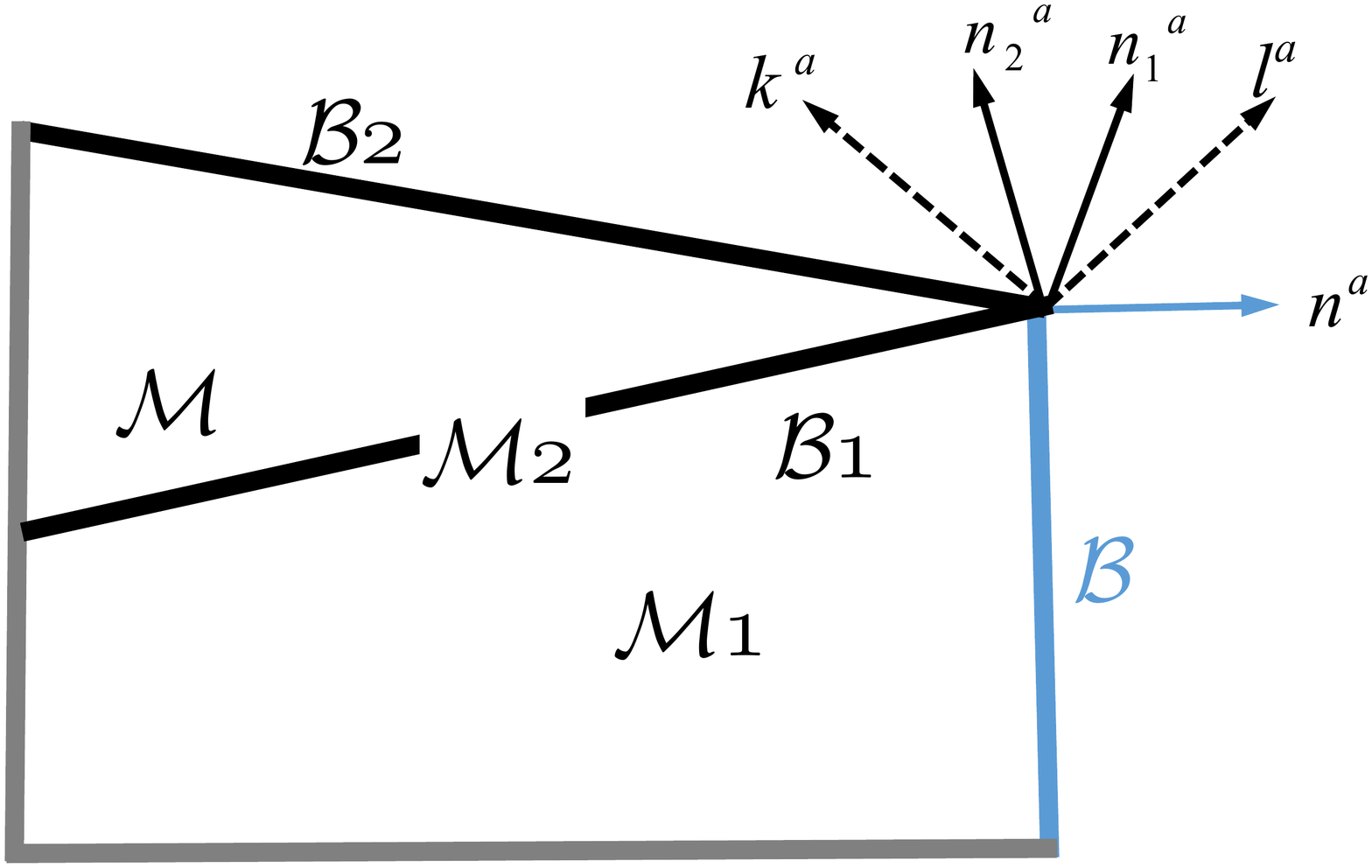}
\label{d}
\end{minipage}
\caption{The corner intersected by $\math{B}_1$ and $\math{B}_2$ can be regarded as the addition or subtraction of two corners by introducing an auxiliary segment $\math{B}$.}
\label{joints}
\end{figure*}
The additivity rule is supposed to be valid not only for the bulk term and surface term, but also for the corner term. With this in mind, one can derive the corner term for any other spacelike joint from the previous one. For instance, regarding the case (a) in \fig{joints}, the corresponding corner term can be obtained as a sum of two corner terms as
\ba\begin{aligned}
I_{\math{C}_a}&=I_{\math{B}\cap\math{B}_1}+I_{\math{B}\cap\math{B}_2}\\
&=-\int_{\math{B}\cap\math{B}_1}\hat{\Y}\h_1dS-\int_{\math{B}\cap\math{B}_2}\hat{\Y}\h_2dS\\
&=-\int_{\math{C}}\hat{\Y}\lf(\h_{1}+\h_{2}\rt) dS\\
&=-\int_{\math{C}}\hat{\Y}\h_a dS\,,
\end{aligned}\ea
where we have introduced an auxiliary segment $\math{B}$.
Note that it follows from \eq{et1} that
\ba
\h_1&=& \ln\lf(l\.n_1\rt)- \ln \lf(-l\. n\rt)\,,\\
\h_2&=& -\ln\lf(l\.n_2\rt)+ \ln \lf(-l\. n\rt)\.
\ea
Thus we have
\ba
\h_a=\ln\lf(l\.n_1\rt)-\ln\lf(l\.n_2\rt)\,.
\ea
Similarly, for the case (b), (c), and (d), the corner term can be readily expressed as minus the Wald entropy density multiplied by the boost parameter with
\ba
\h_b&=&\ln\lf(l\.n_1\rt)-\ln\lf(-l\.n_2\rt)\,,\\
\h_c&=&\ln\lf(l\.n_1\rt)- \ln\lf(-l\.n_2\rt)\,,\\
\h_d&=& \ln\lf(l\.n_2\rt)-\ln\lf(l\.n_1\rt)\,.
\ea

\section{Null segments}\label{se4}
\subsection{Variation of geometric quantities}\label{se41}
We now consider the null segment of the boundary $\math{N}$, which is foliated by an outward-directed null geodesic $k^a=(\frac{\partial}{\partial\lambda})^a$ of a cross section $\math{S}$. We further introduce a null vector field $l_a$ on $\math{N}$, which is normal to $\math{S}$ and satisfies $k^al_a=-1$. With this, the metric can be decomposed as
\ba\label{gkn}
g^{ab}=-k^a l^b-k^b l^a+\s^{ab}\,,
\ea
where $\s^{ab}$ is tangent to $\math{S}$.

In what follows, we shall work with the gauge in which the location of such a null segment as well as its foliation structure keeps unchanged under the variation, $i.e.$,
\ba
\d k_a=\d\a k_a\,,\  \ \ \ \,\d k^a=0\,,
\ea
which implies
\ba
\d l_a=\dbar\b k_a\,,
\ea
where $\d\a=\d g^{ab}k_al_b$ and $\dbar\b=\frac{1}{2}\delta g^{ab}l_al_b$. Furthermore, by $l_al^a=0$ and $k_al^a=-1$, one can obtain
\ba
\d l^a= -\dbar \b\, k^a-\d \a\, l^a +\dbar l^a
\ea
with $\dbar l^a$ tangent to $\math{S}$. Whence the variation of the metric is given by
\ba\label{dgnull}\begin{aligned}
\d g^{ab}&=2\dbar \b\, k^a k^b + \d \a(k^a l^b+k^b l^a) \\
&- k^a \dbar l^b-k^b \dbar l^a+\d \s^{ab}\,.
\end{aligned}\ea

The geodesic equation reads
\ba\label{kappa}
k^a\grad_a k^b=\k\,k^b\,,
\ea
where $\k$ measures the failure of $\l$ to be an affine parameter.
Whence we have the following two expressions for the variation of $\k$ as
\ba
\d \k&=&-\d\lf(l^ak^b\grad_bk_a\rt)=k^a\grad_a\d\a+\d \G^c{}_{ab}l^ak^bk_c\,,\\
\d \k&=&-\d\lf(l_ak^b\grad_bk^a\rt)=-\d \G^a{}_{bc}l_ak^bk^c\,,
\ea
which give rise to
\ba\label{eq1}
\d \G^c{}_{ab}l^ak^bk_c-\d \G^a{}_{bc}l_ak^bk^c=2\d \k-k^a \grad_{a}\d\a\,.
\ea
\subsection{Surface term on the boundary}\label{se42}
For the null segment $\math{N}$, the boundary term in the variation of the bulk action can be expressed as
\ba\begin{aligned}
&\int_{\math{N}}\dbar v^a d\S_a=\int_{\math{N}}k_a\dbar v^a d\l dS\\
&=\int_{\math{N}}\lf[ 2k_c{{\y}_a}{}^{bcd}\d{\G^a}_{bd}+2k_c\d g_{bd}\grad_a{\y}^{abcd}\rt]d\l dS\,.
\end{aligned}\ea
By insertion of \eq{gkn}, we have
\ba\begin{aligned}
&k_c{{\y}_a}{}^{bcd}\d{\G^a}_{bd}=\frac{1}{4}\hat{\Y}\lf(\d\G^a{}_{bc}k_ak^bl^c-\d\G^{c}{}_{ab}k^ak^bl_c\rt)\\
&+k^a \y_{{acdf}} \sigma ^{{bc}} \sigma ^{{ed}} \sigma ^{{hf}} \nabla _h\delta g_{{be}}-k^a k^b \y_{ {bcdf}} l^c \sigma ^{ {ed}} \sigma ^{ {hf}} \nabla _h { \delta  g}_{ {ae}}\\
&+k^a k^b k^c \y_{ {cedf}} l^d l^e \sigma ^{ {fh}} \left(\nabla _h { \delta  g}_{ {ab}}-\nabla _b { \delta  g}_{ {ah}}\right)\\
&+k^a k^b \y_{ {bdcf}} k^c l^d l^e \sigma ^{ {fh}} \left(\nabla _a { \delta  g}_{ {eh}}+\nabla _h { \delta  g}_{ {ae}}-2 \nabla _e { \delta  g}_{ {ah}}\right)\\
&+k^a \y_{ {adbf}} k^b l^c \sigma ^{ {ed}} \sigma ^{ {hf}} \left(\nabla _c { \delta  g}_{ {eh}}-\nabla _h { \delta  g}_{ {ec}}\right)\\
&+k^a k^b \y_{ {bdcf}} l^c \sigma ^{ {ed}} \sigma ^{ {hf}} \left(\nabla _a { \delta  g}_{ {eh}}-\nabla _h { \delta  g}_{ {ea}}\right)\,,
\end{aligned}\ea
where $\hat{\Y}=4\y_{abcd}k^al^bk^cl^d=\y^{abcd}\epsilon_{ab}\epsilon_{cd}$ with the binormal given by $\epsilon_{ab}=(k\wedge l)_{ab}$. Substituting \eq{eq1} and \eq{dgnull} into the above expression and make a straightforward but tedious calculation, one can obtain
\ba\label{nullterm}\begin{aligned}
&\int_{\math{N}}\dbar v^a d\S_a=\int_\math{N} \hat{\Y}\d \k d\l dS\\
&- \int_{\math{N}} \frac{d}{d\l}\lf[\lf(\frac{1}{2}\hat{\Y}\d\a-2k^ak^b\y_{acbd}l^c\dbar l^d\rt) dS\rt] d\l\\
&-2\int_{\math{N}} \lf[\tilde{D}_e\lf(k^ak^bl^c\s^{de} \y_{acbd}\d\a-k^ak^b\s^{ed} \y_{acbd}\dbar l^c\rt)\rt] d\l dS\,,
\end{aligned}\ea
where we have already used the condition $\d \s^{ab}=0$ with $\tilde{D}_a$ the covariant derivative operator on $\math{S}$. Below we shall focus on the case in which $\partial\math{S}=0$. Then the last term in \eq{nullterm} vanishes, which leads to
\ba\label{bdryn}\begin{aligned}
&\int_{\math{N}}\dbar v^a d\S_a=\d\lf[\int_\math{N} \hat{\Y} \k d\l dS\rt]\\
&- \int_{ \partial\math{N}^+} \lf(\frac{1}{2}\hat{\Y}\d\a-2k^ak^b\y_{acbd}l^c\dbar l^d\rt) dS\\
&+\int_{\partial\math{N}^-} \lf(\frac{1}{2}\hat{\Y}\d\a-2k^ak^b\y_{acbd}l^c\dbar l^d\rt) dS
\end{aligned}\ea
where we have used $\d\hat{\Y}=0$. Thus the surface term from the null segment $\math{N}$ is given by
\ba
I_{\text{surf}}=-\int_\math{N} \hat{\Y} \k d\l dS\,.
\ea
On the other hand, if the joint on the boundary is intersected by one null and another non-null segment, the variation of the corner term can be obviously expressed as
\ba\label{corner1}\begin{aligned}
\d I_{\text{corner}}&=\pm \int_{ \math{C}_{s\pm}} \lf(\frac{1}{2}\hat{\Y}\d\a-2k^ak^b\y_{acbd}l^c\dbar l^d\rt) dS\\
&+2\int_{\math{C}_{s\pm}}\lf(\e_s\dbar A_s^a\Y_{s ab}dS_s^b\rt)\,.
\end{aligned}\ea

\subsection{Corner term on the boundary}\label{se43}
\begin{figure}
\centering
\begin{minipage}{0.02\textwidth}
  \ \ \\
  \end{minipage}
\includegraphics[width=0.45\textwidth]{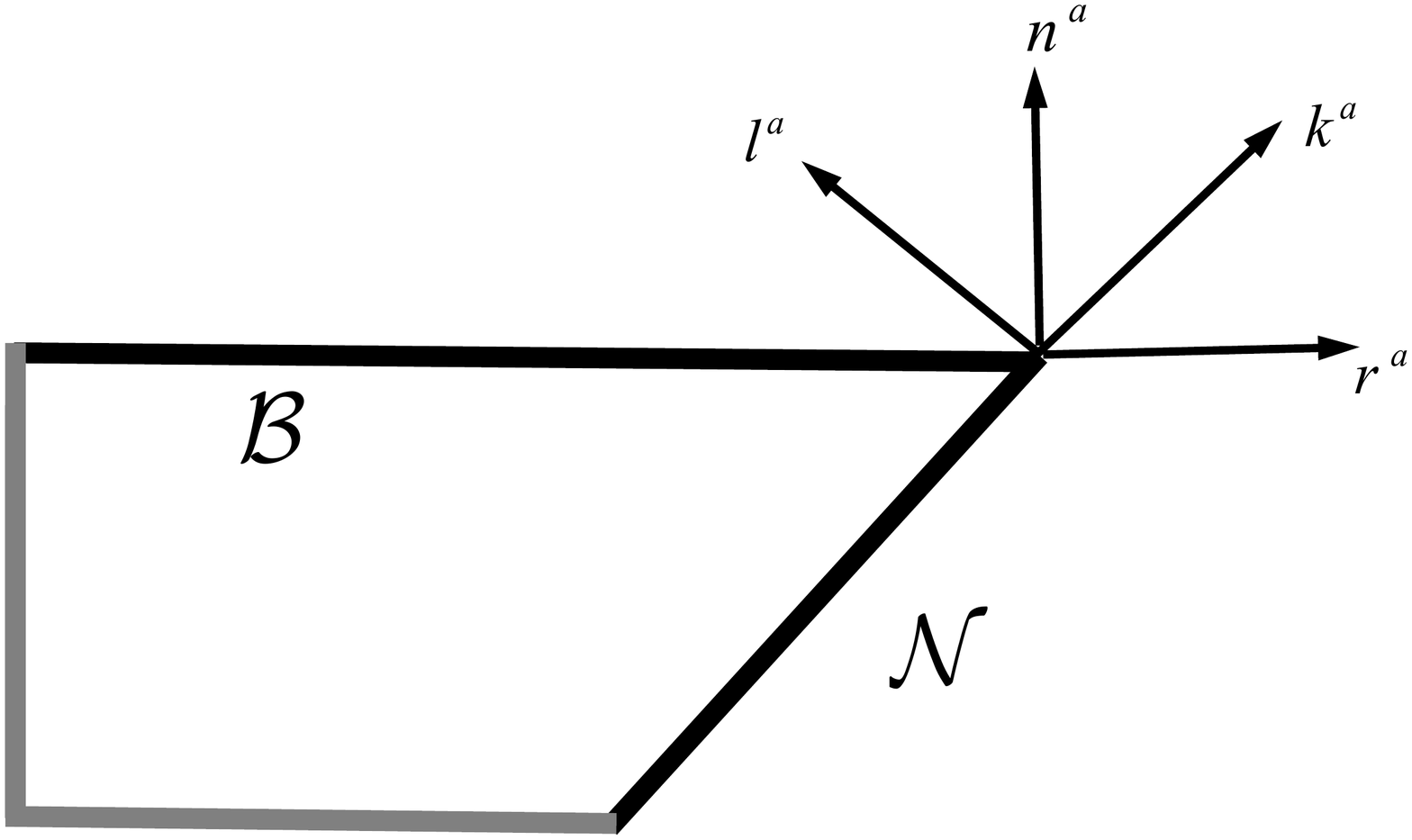}
\caption{The spacelike joint is interjected by a null segment and a spacelike segment.}\label{ns}
\end{figure}
\subsubsection{Joint by a null and a spacelike segment}
As illustrated in \fig{ns}, we first consider the joint which is intersected by a spacelike segment $\math{B}$ and a  null segment $\math{N}$. In this case, there exists a transformation at the joint $\math{C}$, from the pair of normals $\{n^a,r^a\}$ to the double nulls $\{k^a,l^a\}$
\ba\label{trsfkn}\begin{aligned}
k^a&=e^{\chi}\lf(n^a+ r^a\rt)\,,\\
l^a&=\frac{1}{2}e^{-\chi}\lf(n^a-r^a\rt)\,
\end{aligned}\ea
with $\chi$ a scaling factor. Substituting the inverse of this transformation into the following variational identity
 \ba\begin{aligned}
 &\d g^{ab}=2\d a\, n^a n^b-\dbar A^a n^b-\dbar A^b n^a\\
&=2\dbar \b\, k^a k^b + \d \a(k^a l^b+k^b l^a) - k^a \dbar l^b-k^b \dbar l^a
 \end{aligned}\ea
at the joint with $h^{ab}$ and $\s^{ab}$ fixed, we can obtain
\ba\begin{aligned}
&\d g^{ab}=2\dbar \b\, k^a k^b + \d \a(k^a l^b+k^b l^a) - k^a \dbar l^b-k^b \dbar l^a\\
&=\frac{e^{-2\chi}}{2}\lf(\d a-\dbar A^r\rt) k^a k^b+2e^{2\chi}(\d a+\dbar A^r)l^al^b\\
 &+\d a (k^b l^a+k^a l^b)-\frac{e^{-\chi}}{2\sqrt{2}}\lf(k^a\dbar \hat{A}^b+k^b\dbar \hat{A}^a\rt)\\
 &-\sqrt{2}e^{\chi}\lf(l^a\dbar \hat{A}^b+l^b\dbar \hat{A}^a\rt)\,,
\end{aligned}\ea
which gives
\ba\begin{aligned}
\dbar A^r&=-\,\d a\ \ \ \ ,  \ \ \ \dbar \hat{A}^a=\dbar l^a=0\,,\\
\d \a&=\d a\ \ \ \ \ \  ,  \ \ \ \ \dbar \b=\frac{1}{2}e^{-2\chi} \d a\,.
\end{aligned}\ea
Furthermore, by virtue of the variation of $e^{\chi}=-n_ak^a$, one can obtain
\ba
\d a = \d \chi=\d \ln\lf(-n\cdot k\rt)\,.
\ea

With the above preparation, the variation of the corner term can be written as
\ba\begin{aligned}
\d I_{\math{C}}=\int_{\math{C}} \hat{\Y} \d \chi dS
\end{aligned}\ea
which gives the corner term as
\ba\label{J+}
I_{\math{C}}=\int_{\math{C}} \hat{\Y}\,\chi dS
\ea
with
\ba\label{a+}
\chi=\ln\lf(-n\cdot k\rt)\,.
\ea

\subsubsection{Joint intersected by double null segments}\label{se45}
\begin{figure*}
\centering
\includegraphics[width=3in,height=2.1in]{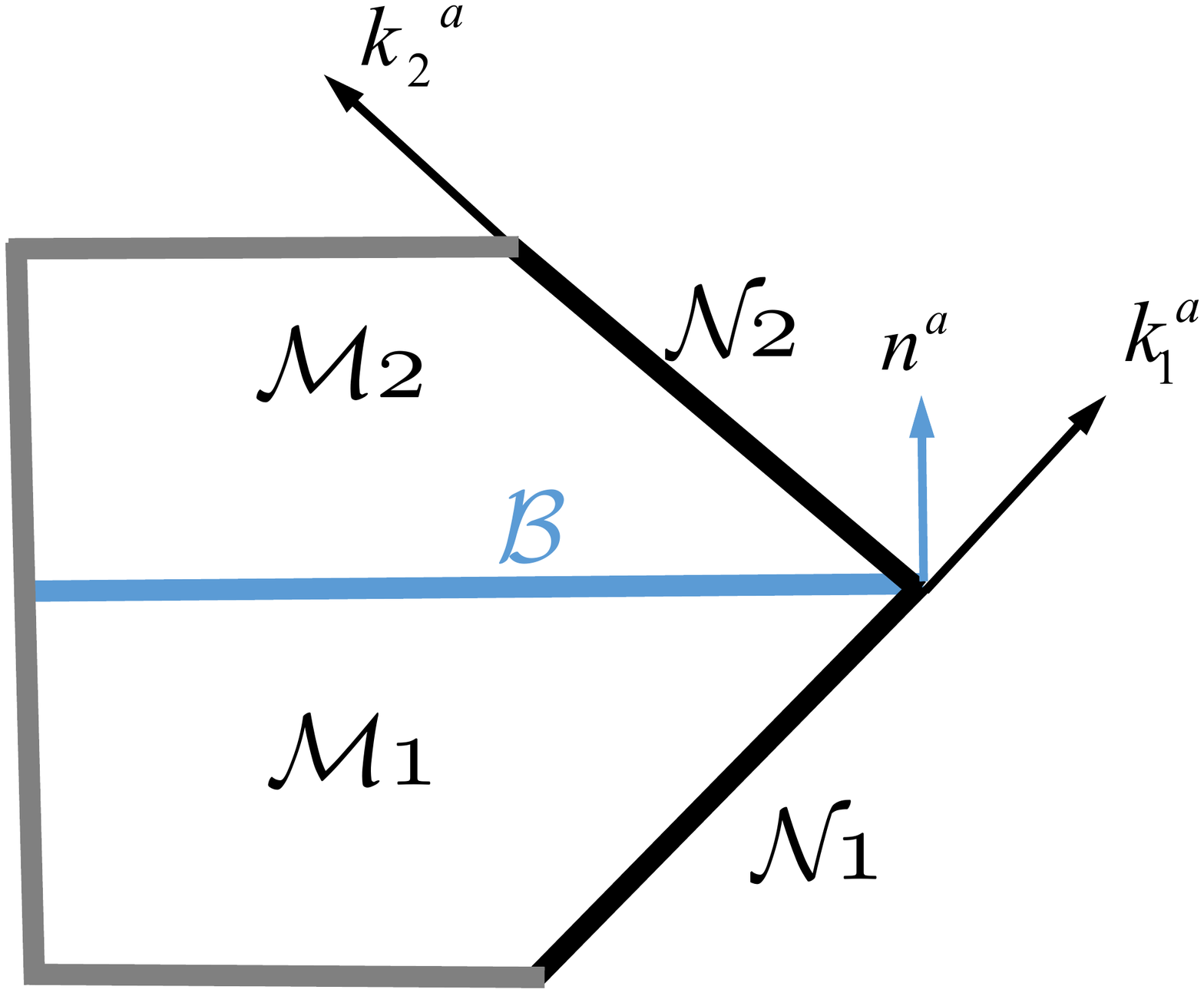}
\includegraphics[width=3in,height=2.1in]{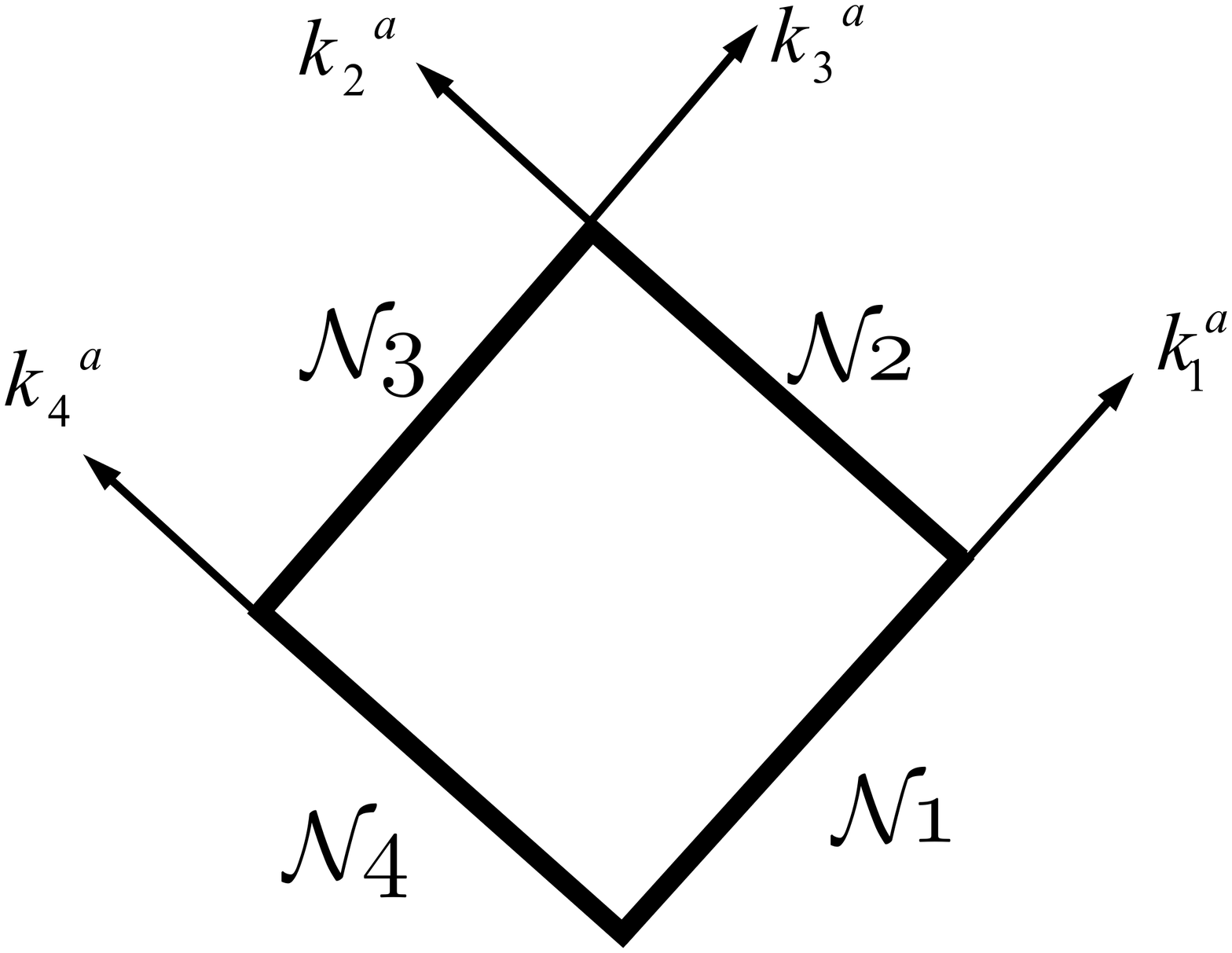}
\caption{The joint is intersected by double null segments can be obtained by the additivity rule.}
\label{ts2}
\end{figure*}
With the corner term obtained before, one can readily derive the corner term for any other type of joint by the additivity rule. As a demonstration and for the later calculations as well, we would like to derive the corner term for the joint intersected by double null segments.
As illustrated in the left panel of \fig{ts2}, we first add an auxiliary spacelike segment $\math{B}$, which divides the corner into two parts. Then by the additivity rule, we have
\ba\begin{aligned}
I_{\math{C}} &= \, \int_{\math{C}}\hat{\Y} \ln\lf(-n\cdot k_1\rt) dS+\, \int_{\math{C}}\hat{\Y}  \ln\lf(-n\cdot k_2\rt) dS\\
&=\int_{\math{C}}\hat{\Y} \lf[ \ln\lf(-n\cdot k_1\rt)+\ln\lf(-n\cdot k_2\rt)\rt] dS\,\\
&=\int_{\math{C}}\hat{\Y}\ln(-\frac{1}{2}k_1\cdot k_2)dS\,.
\end{aligned}\ea
Whence one can readily write down the corner term for the four joints in the right panel of \fig{ts2} as
\ba\label{ncorner}\begin{aligned}
&I_{\text{corner}} =I_{\math{C}_{12}}+I_{\math{C}_{23}}+I_{\math{C}_{34}}+I_{\math{C}_{41}}\\
&=\int_{\math{C}_{12}}\hat{\Y} \,\ln(-\frac{1}{2}k_1\cdot k_2)dS-\int_{\math{C}_{23}}\hat{\Y} \,\ln(-\frac{1}{2}k_2\cdot k_3) dS\\
&+\int_{\math{C}_{34}}\hat{\Y} \,\ln(-\frac{1}{2}k_3\cdot k_4) dS-\int_{\math{C}_{41}}\hat{\Y} \,\ln(-\frac{1}{2}k_4\cdot k_1) dS\,.
\end{aligned}\ea

\subsection{Counter term on the boundary}
Note that the surface term as well as the corner term from the null segment depends on the parametrization of the null generator. In order to eliminate this ambiguity, we can introduce a counter term
\ba
I_{\text{ct}}=-\int_{\math{N}}\hat{\Q}\ln\lf(l_{\text{ct}}\Q\rt)d\l dS\,,
\ea
where $\hat{\Q}=\nabla_a(k^a\hat{\Y})=\frac{1}{\sqrt{\s}}\pd_\l(\hat{\Y}\sqrt{\s})$ and $\Q=\nabla_ak^a=\frac{1}{\sqrt{\s}}\pd_\l(\sqrt{\s})$ is the expansion scalar of the null generator with $l_{ct}$ an arbitrary length scale. To show this, let us consider the reparametrization $\frac{d\bar{\l}}{d\l}=e^{-\b}$, which gives
\ba\begin{aligned}
&\bar{k}^a=e^{\b}k^a\,,\ \ \ \ \bar{\k}=e^{\b}\lf(\k+\pd_\l\b\rt)\,,\\
&\bar{\Q}=e^{\b}\Q\,,\ \ \ \ \  \bar{\hat{\Q}}=e^{\b}\hat{\Q}\,.
\end{aligned}\ea
As a result, we have
\ba\begin{aligned}
&\bar{I}_{\text{surf}}+\bar{I}_{\text{corner}}=I_{\text{surf}}+I_{\text{corner}}\\
&-\int_{\math{N}}\hat{\Y}(\pd_\l\b) d\l dS+\int_{\pd\math{N}^{+}}\hat{\Y}\b dS-\int_{\pd\math{N}^-}\hat{\Y}\b dS\\
        & =I_{\text{surf}}+I_{\text{corner}}+\int_{\math{N}}\frac{\b}{\sqrt{\s}}\pd_\l(\hat{\Y}\sqrt{\s}) d\l dS\,, \\
   \end{aligned}
\ea
and
\ba
\bar{I}_{\text{ct}}=I_{\text{ct}}-\int_{\math{N}}\frac{\b}{\sqrt{\s}}\pd_\l(\hat{\Y}\sqrt{\s}) d\l dS\,,
\ea
which implies
\ba
I_{\text{surf}}+I_{\text{corner}}+I_{\text{ct}}
\ea
is invariant under the above reparametrization.
\section{Application: Case studies for action growth rate}\label{se5}
\subsection{Case 1: SAdS spacetime}\label{case1}
We shall apply our above result to calculate the action growth rate of the WDW patch in the SAdS spacetime for $F(R)$ gravity and critical gravity, respectively. The SAdS metric is obtained originally as a solution to Einstein equation with a negative cosmological constant, $i.e.$,
\ba\label{Rab}
R_{ab}=-\frac{d+1}{L^2}g_{ab}
\ea
with $L$ the AdS curvature radius. Its $(d+2)$-dimensional expression is given by
\ba\label{ds2}
ds^2=-f(r)dt^2+\frac{dr^2}{f(r)}+r^2 d\Omega_{d,k}^2\,,
\ea
where $f(r)=\frac{r^2}{L^2}+k-\frac{\w^{d-1}}{r^{d-1}}$ is the blackening factor, and $k=\{+1,0,-1\}$ denotes the $d$-dimensional spherical, planar, and hyperbolic geometry, individually. The horizon $r=r_h$ lies in the location where $f(r_h)=0$.

\begin{figure}
\centering
\begin{minipage}{0.02\textwidth}
  \ \ \\
  \end{minipage}
\includegraphics[width=0.5\textwidth]{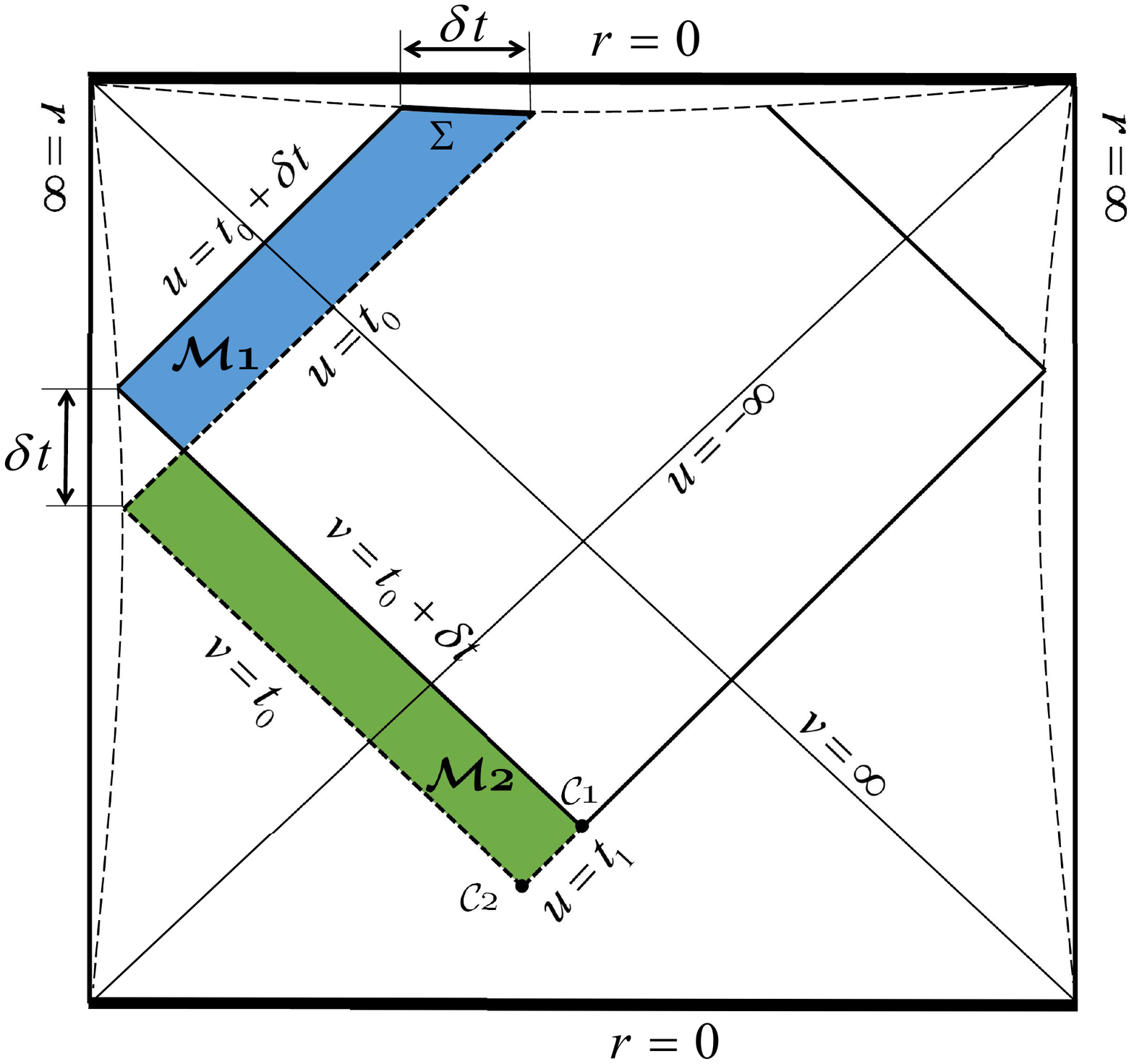}
\caption{ Wheeler-DeWitt patches of a Schwarzschild-AdS black hole.}\label{WdW}
\end{figure}

 As illustrated in the Penrose diagram of the SAdS spacetime \fig{WdW}, $I(t_L,t_R)$, denoted as the action for the WDW patch determined by the time slices on the left and right AdS boundaries, is invariant under the time translation, i.e., $I(t_L+\d t,t_R-\d t)=I(t_L,t_R)$. Thus the action growth can be computed as $\d I = I(t_0+\d t,t_1)-I(t_0,t_1)$, where the time on the right boundary has been fixed. To regulate the divergence near the AdS boundary, a cut-off surface $r=r_{\text{max}}$ is introduced. In addition, we also introduce a spacelike surface $r=r_{\text{min}}$ to avoid running into the spacelike singularity inside of the SAdS black hole. As such, we shall focus on the situation in which the boundary consists solely of null and spacelike segments only with spacelike joints. In addition, for simplicity we shall adopt the affine parameter for the null generator of null segments such that the surface term vanishes for null segments. With this in mind, we have
\ba\begin{aligned}\label{agr}
\d I &=I_{\math{M}_1}-I_{\math{M}_2}+I_{\S}+I_{\math{C}_1}-I_{\math{C}_2}+\d I_{\text{ct}}\,.\\
\end{aligned}\ea
 Here $\math{M}_1$ is bounded by $u=t_0$, $u=t_0+\d t$,$v=t_0+\d t$, and $r=r_\text{min}$. $\math{M}_2$ is bounded by $u=t_0$, $v=t_0$, $v=v_0+\d t$, and $u=t_1$. The null coordinates are defined as $u=t+r^*(r)$ and $v=t-r^*(r)$ with $r^*(r)=\int \frac{dr}{f}$.

\subsubsection{$F(R)$ gravity}\label{se51}
For a general $F(R)$ gravity, the equation of motion reads
\ba
F'(R) R_{ab}-\frac{1}{2}F(R)g_{ab}-(\grad_a \grad_b-g_{ab}\grad^c\grad_c)F'(R)=0\,,\nn\\
\ea
and the auxiliary field as well as its decedents can be expressed as
\ba\begin{aligned}
\y^{abcd}&=\frac{1}{2}\lf(g^{ac}g^{bd}-g^{ad}g^{bc}\rt)F'(R)\,,\\
\Y_{ab}&=-\frac{1}{2}h_{ab}F'(R)\,,\\
\hat{\Y}&=-2F'(R)\,.
\end{aligned}\ea
Whence the full on-shell action can be simplified as
\ba\label{ac2}\begin{aligned}
&I=I_{\text{bulk}}+I_{\text{surf}}+I_{\text{corner}}+I_{\text{ct}}\\
&=\int_{\math{M}}d^{d+2}x\sqrt{-g}F(R)-2\sum_{s}\lf(\int_{\math{B}_s}K F'(R) d\S\rt)\\
&-2 (-1)^\l \int_{\math{C}_{\l}} \,c_{\l}F'(R) dS-\int_{\math{N}}\hat{\Q}\ln\lf(l_{\text{ct}}\Q\rt)d\l dS\,.
\end{aligned}\ea

In what follows, we shall consider the special case, in which there exists an $R_0$ such that
\ba
F(R_0)=\frac{2R_0}{d+2}F'(R_0)\,,
\ea
where the prime denotes the derivative with respect to $R$. As such, (\ref{Rab}) with $L^2=-\frac{(d+1)(d+2)}{R_0}$ satisfies the above equation of motion. Accordingly, the SAdS metric can be regarded as its solution.

With the above preparation, now let us use (\ref{agr}) to calculate the action growth rate in our $F(R)$ gravity. So we only need to keep the first order of $\d t$ below for each term in (\ref{agr}).  First, with the $(u,r)$ coordinates, we have
\ba\label{M1}\begin{aligned}
I_{\math{M}_1}&=F(R_0)\W_{d,k}\int_{t_{0}}^{t_0+\d t}du \int_{r_\text{min}}^{\r(u)}r^d dr\\
&=\frac{F(R_0)\W_{d,k}}{d+1}r_{\text{max}}^{d+1}\d t\,,
\end{aligned}\ea
where $r=\r(u)$ is the solution to the equation $v(u,r)=t_0+\d t$ and $r_{\text{min}}$ has been set to zero in the end. Similarly, with the $(v,r)$ coordinates, we have
\ba\label{M2}\begin{aligned}
I_{\math{M}_2}&=F(R_0)\W_{d,k}\int_{t_{0}}^{t_0+\d t}dv \int_{\r_1(v)}^{\r_0(v)}r^d dr\\
&=\frac{F(R_0)\W_{d,k}}{d+1}\lf(r_{\text{max}}^{d+1}-r_1^{d+1}\rt)\d t\,,
\end{aligned}
\ea
where $r=\r_0(v)$ is the solution to the equation $u(v,r)=t_0$ and $r=\r_1(v)$ is the solution to the equation $u(v,r)=t_1$ with $r_1$ the $r$ coordinate of $\math{C}_1$. Thus the difference between $I_{\math{M}_1}$ and $I_{\math{M}_2}$
 \ba
 I_{\math{M}_1}-I_{\math{M}_2}=\frac{F(R_0)\W_{d,k}}{d+1}r_1^{d+1}\d t\,.
 \ea
For the surface term, we have
\ba\begin{aligned}
I_{\S}&=-2F'(R_0)\int_{\math{S}}K d\S \\
&= (d+1)F'(R_0)\W_{d,k}\w^{d-1}\d t,
\end{aligned}\ea
where we have used the expression $K=-\frac{1}{r^d}\frac{d}{dr}\lf(r^d\sqrt{-f}\rt)$ for the spacelike surface $r=r_{\text{min}}$ and let $r_{\text{min}}\to 0$ in the end.

In order to write down the explicit expression for the difference between the two corner terms from $\math{C}_1$ and $\math{C}_2$, we shall choose $k_{1a}=\grad_a u$ and $k_{2a}=-\grad_a v$. Note that $k_1\cdot k_2=-\frac{2}{f}$. Then by (\ref{ncorner}), we can obtain
\ba\begin{aligned}
&I_{\math{C}_1}-I_{\math{C}_2}=F'(R_0)\W_{d,k}\\
&\times\lf[r_1^d f'(r_1)+d \,r_1^{d-1}f(r_1)\ln\lf(-f(r_1)\rt)\rt]\d t\,,
\end{aligned}\ea
where we have used
\ba\label{dr}
\d r=r_1-r_2=-\frac{1}{2}f(r_1)\d t\,
\ea
with $r_2$ the $r$ coordinate of $\math{C}_2$.

In the SAdS spacetime, the counter term can be expressed as
\ba
I_\text{ct}=2F'(R_0)\int_{\math{N}}\Q\ln(l_\text{ct}\Q)d\l dS\,.
\ea
By the translation symmetry, there are only two null segments contributing to the action growth. The first one comes from the null segment $u=t_1$ with $r$ as the affine parameter, $i.e.$, $k_1^{a}=\lf(\frac{\pd}{\pd r}\rt)^a$, which gives rise to the expansion $\Q=\frac{d}{r}$. As a result, the corresponding counter term can be written as
\ba\label{Sct1}\begin{aligned}
&I_\text{ct}^{(1)}=2d\W_{d,k}F'(R_0)\int_{r_2}^{r_\text{max}}dr\,r^{d-1}\ln\left(\frac{d l_\text{ct}}{r}\right)\\
&=2\W_{d,k}F'(R_0)\times\\
&\lf[r_{\text{max}}^d\ln\lf(\frac{d l_\text{ct}}{r_{\text{max}}}\rt)-r_2^d\ln\lf(\frac{d l_\text{ct}}{r_2}\rt)
+\frac{1}{d}\lf(r_\text{max}^d-r_2^d\rt)\rt]\,.
\end{aligned}
\ea
Obviously, as to the counter term from the second null segment $v=t_0$, we have $I_\text{ct}^{(2)}=I_\text{ct}^{(1)}$. By \eq{dr}, the growth of the counter term can be written as
\ba\begin{aligned}
\d I_{\text{ct}}=2d\W_{d,k}F'(R_0) f(r_1)r_1^{d-1}\ln\lf(\frac{d l_\text{ct}}{r_1}\rt)\d t\,.
\end{aligned}\ea
 Then summing all the previous terms, we end up with
\ba\begin{aligned}
\d I&=\W_{d,k}F'(R_0)\lf[-2\frac{r_1^{d+1}}{L^2}+(d+1)\w^{d-1}\right.\\
&\left.+r_1^d f'(r_1)+d \,r_1^{d-1}f(r_1)\ln\lf(\frac{-f(r_1)d^2l_\text{ct}^2}{r_1^2}\rt)\rt]\d t\\
&=2d\,\W_{d,k}F'(R_0)\w^{d-1}\\
&\times\lf[1+\frac{1}{2}\lf(\frac{r_1}{\w}\rt)^{d-1}f(r_1)\ln\lf(\frac{-f(r_1)d^2l_\text{ct}^2}{r_1^2}\rt)\rt]\d t\\
&= 2M_F \lf[1+\frac{1}{2}\lf(\frac{r_1}{\w}\rt)^{d-1}f(r_1)\ln\lf(\frac{-f(r_1)d^2l_\text{ct}^2}{r_1^2}\rt)\rt]\d t\,,
\end{aligned}\ea
where
\ba\label{mass1}
M_F=\W_{d,k}d\w^{d-1} F'(R_0)\,
\ea
is the ADM mass\cite{DKH}. As a result, the action growth rate is given by
\ba
\dot{I}=2 M_F \lf[1+\frac{1}{2}\lf(\frac{r_1}{\w}\rt)^{d-1}f(r_1)\ln\lf(\frac{-f(r_1)d^2l_\text{ct}^2}{r_1^2}\rt)\rt]\,,\nn\\
\ea
which reduces to
\ba
\dot{I}=2 M_F\,,
\ea
in the late time limit with $r_1\to r_h$. It is noteworthy that this late time behavior is also obtained by different approaches in \cite{AFNV,GWLL,WYY}.

\subsubsection{Critical gravity}
Now let us move onto the critical gravity. The original bulk action is given by\cite{AFNV,LP},
\ba\label{Ibulk}\begin{aligned}
&I_{\text{bulk}}=\int_{\math{M}}d^{d+2}x\sqrt{-g}\\
&\times\lf[R-2\L-\frac{1}{m^2}\lf(R^{ab}R_{ab}-\frac{d+2}{4(d+1)}R^2\rt)\rt]\,,
\end{aligned}\ea
where $m$ is a dimensionful parameter. Whence the corresponding equation of motion can be obtained as
\ba
&\left[8 \Lambda  m^2+8 d \Lambda  m^2+4 (d+1) \left(R_{ {cd}} R^{ {cd}}-m^2 R\right)\right.\nn\\
&\left.+ (d+2)\left(4\nabla _c\nabla ^cR-R^2\right)-8  (d+1)\grad_{c}\grad_dR^{ {cd}}\right]g_{ {ab}}\nn\\
&+4 \left[ R_{ {ab}} \left(2 (d+1) m^2+(d+2) R\right)-4 (d+1) R_{ {bc}} R^c{}_a\right]\nn\\
&+4 \left[-(d+2) \nabla _a\nabla _bR +2(d+1)\grad_a\grad_cR^c{}_b \right.\nn\\
&\left.+2(d+1)\grad_b\grad_cR^c{}_a{}-2 (d+1)\grad_c\grad^cR_{ {ab}}\right]=0\,,\nn\\
\ea
and the auxiliary field as well as its decedents reads
\ba
&&\y^{abcd}=\lf(1+\frac{2+d}{2(1+d)m^2}R\rt)g^{c[a}g^{b]d}\nn\\
&&-\frac{1}{m^2}R^{a[c}g^{d]b}-\frac{1}{m^2}R^{c[b}g^{a]d}\,,\nn\\
&&\Y_{ab}=-\frac{1}{2}\lf(1+\frac{2+d}{2(1+d)m^2}R\rt)h_{ab}\\
&&-\frac{1}{2m^2}\lf[R_{cd}n^cn^dg_{ab}-R_{ab}-n^c\lf(n_a R_{bc}+n_b R_{ac}\rt)\rt]\,,\label{Y1}\nn\\
&&\hat{\Y}=-2+\frac{1}{m^2}\lf(2R_{ab}r^a r^b-2R_{ab}n^an^b-\frac{d+2}{d+1}R\rt)\label{P1}\,.\nn
\ea
It is not hard to show that with
\ba
\L=\frac{d (d+1) \left(d^2-2 d-4 L^2 m^2\right)}{8 L^4 m^2}\,,
\ea
\eq{Rab} satisfies the equation of motion. So in this case, the SAdS metric is also the solution to the critical gravity. With this solution, one can obtain
\ba\begin{aligned}
&I_{\text{bulk}}=-2\frac{d+1}{L^2}\lf(1-\frac{d^2}{2L^2m^2}\rt)\int_{\math{M}}d^{d+2}x\sqrt{-g}\,,\\
&\Y_{ab}=-\frac{1}{2}\lf(1-\frac{d^2}{2L^2m^2}\rt)h_{ab}\,,\\
&\hat{\Y}=-2\lf(1-\frac{d^2}{2L^2m^2}\rt)\,.
\end{aligned}\ea
Following the same calculation as $F(R)$ gravity, one can easily obtain the action growth rate for the critical gravity as
\ba
\dot{I}=2 M_C \lf[1+\frac{1}{2}\lf(\frac{r_1}{\w}\rt)^{d-1}f(r_1)\ln\lf(\frac{-f(r_1)d^2l_\text{ct}^2}{r_1^2}\rt)\rt]\,,\nn\\
\ea
where
\ba\label{mass2}\begin{aligned}
M_C=d\, \W_{d,k}\w^{d-1}\lf(1-\frac{d^2}{2L^2m^2}\rt)\,
\end{aligned}\ea
is the ADM mass for the critical gravity\cite{LP}. The late time action growth rate is the same as that obtained in \cite{AFNV} by using the approach developed in \cite{BRSSZ1,BRSSZ2}.

\subsection{Case 2: The asymptotically AdS black hole for the critical Einsteinian cubic gravity}
In this subsection, we consider the $4$-dimensional the critical Einsteinian cubic gravity. The corresponding bulk action is given by\cite{BC}
\ba\begin{aligned}
I_\text{bulk}&=\int_\math{M}d^4x\sqrt{-g}F\\
&=\int_\math{M}d^4x\sqrt{-g}\lf(R-2\L +\l \math{P}\rt)\,,
\end{aligned}\ea
where the cubic invariant polynomial term $\math{P}$ of the Riemann tensor reads
\ba\begin{aligned}
\math{P}&=12 R_{acbd}R^{cedf}R_e{}^a{}_f{}^b+R_{ab}{}^{cd}R_{cd}{}^{ef}R_{ef}{}^{ab}\\
&-12R_{abcd}R^{ac}R^{bd}+8R_a{}^bR_b{}^cR^a{}_c
\end{aligned}\ea
with $\lambda$ the coupling constant.

In terms of the auxiliary field
\ba\begin{aligned}
&\y_{abcd}=\frac{1}{2}\lf(g_{ac}g_{bd}-g_{ad}g_{bc}\rt)+6\l \left(R_{ad}R_{bc}\right.\\
&-R_{ac}R_{bd}+g_{bd}R_a{}^eR_{ce}-g_{ad}R_b{}^eR_{ce}\\
&-g_{bc}R_a{}^eR_{de}
+g_{ac}R_b{}^eR_{de}-g_{bd}R^{ef}R_{aecf}\\
&+g_{bc}R^{ef}R_{aedf}+g_{ad}R^{ef}R_{becf}-g_{ac}R^{ef}R_{bedf}\\
&-3R_a{}^e{}_d{}^fR_{becf}
+3R_a{}^e{}_c{}^fR_{bedf}+\frac{1}{2}R_{ab}{}^{ef}R_{cdef})\,,
\end{aligned}\ea
the equation of motion can be expressed as
\ba\begin{aligned}
\y_{acde}R_b{}^{cde}-\frac{1}{2}g_{ab}F-2\grad^c\grad^d\y_{acdb}=0\,.
\end{aligned}\ea
As shown in \cite{FHML}, when the parameters satisfies the following critical relation
\ba
\L=-\frac{2}{L^2}\,,\ \ \ \ \l=-\frac{L^4}{24}\,,
\ea
the above equation of motion admits a static asymptotically AdS black hole solution, whose line element can be written as
\ba\label{ds2c}
ds^2=-f(r)dt^2+\frac{dr^2}{f(r)}+r^2d\Omega_{2,k}^2
\ea
with the blackening factor
\ba
f(r)=\frac{r^2}{L^2}+k-\m\,.
\ea
Then following the exact same procedure, one can obtain
\ba
I_{\math{M}_1}&-I_{\math{M}_2}=4\W_{2,k}r_1\lf(\m-\frac{r_1^2}{L^2}\rt)\d t\,.
\ea
By using \eq{Isurf}, one can further find that the surface term of $\S$ vanishes. In addition, the straightforward calculation gives rise to the following corner term
\ba\begin{aligned}
&I_{\math{C}_1}-I_{\math{C}_2}=4\W_{2,k}r_1\\ &\times\lf[\lf(\frac{r_1^2}{L^2}-\m\rt)+f(r_1)\ln\lf(-f(r_1)\rt)\rt]\d t\,.
\end{aligned}\ea
At last, the counter term contribution of the null segments can be obtained as
\ba
\d I_\text{ct}=8\W_{2,k} r_1 f(r_1)\ln\lf(\frac{2l_\text{ct}}{r_1}\rt)\d t\,,
\ea
where we have used $\hat{\Q}=-\frac{8}{r}$.
By summing all the previous terms, we end up with
\ba\begin{aligned}
\dot{I}=4\W_{2,k}r_1f(r_1)\ln\lf(-\frac{4f(r_1)l_\text{ct}^2}{r_1^2}\rt)\,.
\end{aligned}\ea
In the late time limit, the action growth rate apparently vanishes. However, this late time behavior still saturates the Lloyd bound because the mass of this black hole also vanishes\cite{FHML}.

\section{Conclusion}\label{se6}

We have presented a complete discussion of the variational problem for $F($Riemann$)$ gravity with a non-smooth boundary. In order to give rise to a well posed variational principle, we must supplement the surface term and corner term to the bulk action. Following the method developed in \cite{LMPS}, we obtain a general formula for the boundary term, where the corner term can be obtained by integrating the Wald entropy density weighted by a transformation parameter between the two intersected segments. When the involved segment is null, we are also required to add a counter term to make the full boundary term invariant under the reparametrization.

Then motivated by the CA conjecture, we apply the resulting full action to evaluate the full time action growth rate of the WDW patch in the SAdS spacetime for the $F(R)$ gravity and critical gravity, as well as in an asympotically AdS black hole for the critical Einsteinian cubic gravity. For the $F(R)$ and critical gravity,  the late time action growth rate shares exactly the same behavior as those obtained by other approaches. For the critical Einsteinian cubic gravity, we find that the late time action growth rate vanishes but still saturates the Lloyd bound.

\begin{acknowledgments}
J.J. is partially supported by NSFC with Grant No.11375026, 11675015, and 11775022. H.Z. is supported in part by FWO-Vlaanderen through the project G020714N, G044016N, and G006918N. He is also an individual FWO Fellow supported by 12G3515N.
\end{acknowledgments}

\end{document}